\documentclass[twocolumn,showpacs,preprintnumbers,amsmath,amssymb]{revtex4}

\begin{document}
\title{Quasi-one-dimensional disordered systems: fluctuations, transport 
 and interplay}
\author{A.V. Plyukhin}
 \affiliation{ Department of Physics and Engineering Physics,
 University of Saskatchewan, Saskatoon, SK S7N 5E2, Canada 
}

\date{\today}

\begin{abstract}
In a one dimensional lattice thermal fluctuations destroy the long-range order
making  particles of the lattice  move 
on a scale much larger than the lattice spacing. 
We discuss the assumption that this motion may be responsible
for the transport of localized electrons in a system of weakly coupled chains.
The model with diffusing localization sites gives a temperature-independent 
mobility with a crossover to an activated dependence at high temperature.
This prediction is consistent with and might account for 
experimental results on discotic liquid crystals and certain biopolymers.
\end{abstract}

\pacs{72.20.-i, 72.70.+m, 72.80.Le}

\maketitle



\section{Introduction} 

In one  and two dimensional (D) systems
positional correlations diverge with the system size   
due to thermal fluctuations.
Such  loss of long-range order, known as Landau-Peierls instability, implies 
the divergence of  
the  mean square displacement (MSD) of system's structural units. 
For harmonic lattices this result may be illustrated  by direct 
calculations.  
Let $q_i$ be a displacement of the $i$-th atoms in a harmonic chain from
its equilibrium position $x_i\sim ia$ where $a$ is the lattice spacing. 
Assuming that the atoms are distributed as in thermal equilibrium,
one can show that the atomic MSD $q_i$ is proportional
to the atom's distance from the chain's end,
\begin{equation}
\langle q_i^2\rangle =\frac{k_BT}{m\omega_0^2}i.
\label{eq1}
\end{equation}
Here $\omega_0=\sqrt{k/m}$, $k$ 
is the harmonic force constant, and $m$ is the mass of an atom.
In  a sufficiently  long 
chain the atomic displacement from equilibrium position 
$\langle q_i^2\rangle^{1/2}$ may 
be significantly longer than the equilibrium lattice spacing $a$.
For instance,
for the force constant $k\sim 1 \, N/m$, temperature 
$T\sim 10^2\,K$, and $i=10^4$ the equation (1) predict  
the displacement of order $100\,\AA$.
In the limit of the infinite chain the MSD  diverges
and atomic motion is unbounded. 
The loss of long-range order 
due to Landau-Peierls instability has been observed in X-ray and neutron 
scattering experiments in many quasi-1D and quasi-2D systems such as liquid 
crystals and membranes. 

The divergence of the MSD does not mean, of course, 
that the chain is unstable. While the long range order is lost,
the short order is preserved. Indeed, the result for 
the relative displacement of two atoms reads
\begin{equation}
\langle (q_i-q_{i+j})^2\rangle =\frac{k_BT}{m\omega_0^2}j.
\label{eq2}
\end{equation}
According to this equation the distance between two adjacent atoms ($j=1$) 
does not depend on the chain's length and deviates from $a$ 
by a value of order $(k_BT/m\omega_0^2)^{1/2}$, which is normally small.

In recent paper~\cite{hh} it was suggested that since  individual atoms 
in a 1D lattice are to some extent  delocalized,  
they  may serve as  temporary vehicles for  localized electrons.     
Consider a system of parallel chains separated by the distance 
$b$ which is larger than the lattice spacing $a$.
As will be discussed in the following sections, 
on a long time scale ($\omega_0t\gg 1$) a tagged atom  
in a long isolated harmonic chain behaves as a Brownian particle with
the diffusion constant
\begin{equation}
D_a=\frac{kT}{2m\omega_0},
\end{equation}
which increases linearly with temperature.
On  the other hand, the diffusion constant $D_h$ 
for  electronic  hopping in  a static disordered chain
depends  on temperature exponentially and may be smaller than 
$D_a$ for sufficiently low $T$.
For instance, for the nearest-neighbor hopping over uncorrelated sites  
with the  Gaussian energy distribution 
\begin{equation}
g(\epsilon)=c\exp[-\epsilon^2/2\epsilon_0^2],
\end{equation}
the hopping diffusion constant has the form~\cite{bass}
\begin{equation}
D_h=
\frac{a^2}{2}\,\nu\, \exp\left[
-a/L-\left(\epsilon_0/k_BT\right)^2
\right]
\label{Dh}
\end{equation}
where $L$ is the localization length
of the carrier wave function,
and $\nu$ is the  ``attempt frequency''.
For the typical parameter set $k\sim 10\, N/m$, $\epsilon_0\sim 0.1\, eV$, 
$a\sim 10\,\AA$, $\nu/\omega_0\sim 1$, and $\exp(-a/L)\sim 10^{-3}$,   
one finds that
$D_h\ll D_a$ when  $\epsilon_0/k_BT>1$. 
This estimation 
suggests what we call the hitchhiking mechanism of electronic transport: 
for sufficiently low temperature
the hopping mechanism is responsible only for the transport 
perpendicular to the chains, while 
along the chains electrons are transported predominantly by mobile 
localization sites.

To estimate the diffusion constant $D_\parallel$
for the lateral 
transport along the chains due to the hitchhiking mechanism 
let us assume that the localization is strong 
and each atom is associated with a mobile localization site whose
MSD as a function of time $\langle \Delta q^2(t)\rangle=\langle [q(t)-q(0)]^2\rangle$ is given.
Suppose  also that the transition rates for the inter-chain hopping
do not strongly fluctuate
around a typical value $W_\perp$. Then
the lateral motion of an electron can be considered as 
a 1D random walk with the the time step $\tau\sim1/W_\perp$ and the length step
$l\sim\langle \Delta q^2(\tau)\rangle^{1/2}$. The corresponding diffusion constant is
$l^2/2\tau$, which gives 
\begin{equation}
D_\parallel=\frac{1}{2} W_\perp \langle \Delta q^2(W_\perp^{-1})\rangle.
\label{Destim}
\end{equation}
If the atomic MSD is diffusive 
$\langle q^2(t)\rangle\sim 2D_at$, as in an isolated chain, then 
$D_\parallel$ does not depend on $W_\perp$ and  coincides with the atomic
diffusion coefficient
\begin{equation}
D_\parallel= D_a=\frac{k_BT}{2m\omega_0}.
\label{D}
\end{equation}
The corresponding mobility 
is temperature independent
\begin{equation}
\mu=\frac{eD_\parallel}{k_BT}=\frac{e}{2m\omega_0}=\frac{e\,\omega_0}{2k}.
\label{mmu}
\end{equation}
For $\omega_0=10^{13}\, s^{-1}$ and $k=10\,N/m$ this equation  gives
 $\mu\sim 10^{-3}\,cm^2(Vs)^{-1}$, which is
consistent with  experimental values for 
columnar liquid crystals and 
certain conjugated polymers.

Temperature independent mobility was  observed in many low-dimensional 
soft matter systems, in particular  in discotic 
liquid crystals~\cite{LC} and DNA~\cite{dna}. 
It is usually explained in terms of the polaron model~\cite{LC}
or the dynamical disorder models~\cite{Dyn}. 
The model of hitchhiking  transport gives an alternative 
and very simple explanation, which does not involve any adjustable parameters.
The model  also predicts a crossover to an 
activated  temperature dependence for sufficiently high temperature
when  the inequality $D_h\ll D_a$ is no longer valid, and the conventional
hopping mechanism begins to dominate. Such crossover was reported
for charge transport in DNA~\cite{dna}.

The underlying issue of the model  is the one of delocalized single-particle 
dynamics in low-dimensional lattices. 
The purpose of this paper is to illustrate  this issue for  
two simplest approximations, one of non-interacting chains, 
and the other  of dissipative chains, subjected to the noise and 
friction forces.

\section{Lack of long-range order}
Although delocalization of an atom in low-dimensional lattices is a well-known 
result, it might appear counter-intuitive and is often misinterpreted.
For instance, in~\cite{Lee} the the divergence of the atomic MSD in a 1D chain
is considered as an artifact which arises from the zero-frequency mode 
in the chain with periodic boundary conditions. Actually, delocalization is 
the general property of low-dimensional lattices and does not depend on 
the type of boundary conditions. The question was comprehensively studied 
by Montroll~\cite{Mont}, but the generality of his approach  makes 
it rather complicated. In this section we give a simple derivation 
of Eq. (1) for a harmonic chain and generalize it for the quantum case.

Consider a harmonic chain of $N+2$ atoms with fixed ends. Labeling atoms by
index $i=0,1,...N+1$ one can write the Hamiltonian in the form 
 \begin{equation}
H=\frac{1}{2m}\displaystyle\sum_{i=1}^N{p_i}^2+
\frac{m\omega_0^2}{2}\displaystyle\sum_{i=1}^{N+1}\left(q_i-q_{i-1}\right)^2
\label{H}
\end{equation}
assuming that displacements for the terminal atoms are zero,  
$q_0=q_{N+1}=0$.  
The Hamiltonian  can be diagonalized by means of a normal mode transformation,
\begin{equation}
q_i=\frac{1}{\sqrt{m}}\sum_{j=1}^NA_{ij}Q_j, \,\,\,\,
p_i=\sqrt{m}\sum_{j=1}^NA_{ij}P_j
\label{transformation}
\end{equation}
with normalized eigenvectors
\begin{equation}
A_{ij}=
\left(\frac{2}{N+1}\right)^{\frac{1}{2}}\sin\left(\frac{\pi ij}{N+1}\right)
\end{equation}
which satisfy the orthogonality condition $\sum_{i=1}^N A_{ij}A_{ik}=\delta_{jk}$.
In terms of  normal  coordinates the Hamiltonian assumes the form,
\begin{equation}
H=\frac{1}{2}\displaystyle\sum_{j=1}^N\left({P_j}^2+
{\omega_j}^2{Q_j}^2\right)
\end{equation}
where the  normal mode frequencies are
\begin{equation}
\omega_j=2\omega_0\sin\left[\frac{\pi j}{2(N+1)}\right].
\end{equation}
Then the  atomic MSD can be written as
\begin{equation}
\langle {q_i}^2\rangle=\frac{1}{m}\displaystyle\sum_{jj^\prime}A_{ij}A_{ij'}
\langle Q_jQ_{j^\prime}\rangle
\label{q2}
\end{equation}
where the  average is taken with respect to the canonical 
distribution $\rho=Z^{-1}e^{-H/k_BT}$. Since
\begin{equation}
\langle Q_j Q_{j'}\rangle=\delta_{jj'}\frac{k_BT}{\omega_j^2},\,\,\,
\langle P_j P_{j'}\rangle=\delta_{jj'}k_BT,
\end{equation}
the equation (\ref{q2}) takes the form
\begin{equation}
\langle{q_i}^2\rangle
=\frac{k_BT}{m}\sum_{j=1}^N \frac{{A_{ij}}^2}{{\omega_j}^2}. 
\label{aux}
\end{equation}
Inserting the expressions for $A_{ij}$ and $\omega_j$,
and introducing the new variable 
\begin{equation}
\theta_j=\frac{\pi j}{2(N+1)}
\end{equation}
one obtains 
\begin{equation}
\langle{q_i}^2\rangle=
\frac{k_BT}{4m\omega_0^2}\frac{2}{N+1}\displaystyle\sum_{j=1}^N 
\frac{\sin ^2{\left(2i\theta_j\right)}}{\sin ^2{\theta_j}}.
\end{equation}
Since  $\Delta\theta_j=\theta_{j+1}-\theta_j=\pi/2(N+1)$,
the sum in the above expression can be converted in the limit $N\to\infty$
into the integral as follows
\begin{equation}
\langle{q_i}^2\rangle=\frac{k_BT}{m\omega_0^2}\,
\frac{1}{\pi}\int_0^\frac{\pi}{2} 
\frac{\sin^2{\left(2i\theta\right)}}{\sin ^2{\theta}}d\theta\\
\end{equation}
which eventually gives 
\begin{equation}
\langle{q_i}^2\rangle=\frac{k_BT}{m\omega_0^2}i=\frac{k_BT}{k}i.
\label{msd}
\end{equation}
The equation (2) can be derived in a similar way.

In fact, the  result (\ref{msd}) can be obtain without the normal mode transformation,
but using instead new coordinates $\delta_i=q_i-q_{i-1}$ 
and noticing that 
$q_i=\sum_{k=1}^{i}\delta_k$.
This formula is the manifestation of cumulative nature of deviation
from equilibrium in 1D systems. 
Another comment concerns the equations (\ref{aux}), the summation of which 
gives 
\begin{equation}
\sum_i\langle q_i^2\rangle=\frac{k_BT}{m}\sum_j\omega_j^{-2}.
\end{equation}
This relation does not involve $A_{ij}$ and is believed to be a
general result~\cite{Go}.

The shortest way to generalized the result (\ref{msd}) for the quantum case
is to use the quantum mechanical formula for the average 
energy of the oscillator, corresponding to a $j$-th mode
\begin{equation}
\langle E_j\rangle=
\frac{\hbar\omega_j}{2}\coth{\left(\frac{\hbar\omega_j}{2k_BT}\right)}.
\end{equation}
Since the average potential and kinetic energies are equal, 
\begin{equation}
\frac{\omega_j^2}{2}\langle Q_j^2\rangle=
\frac{1}{2}\langle P_j^2\rangle=
\frac{1}{2}\langle E_j\rangle=
\frac{\hbar\omega_j}{4}\coth{\left(\frac{\hbar\omega_j}{2k_BT}\right)}
\nonumber
\end{equation}
one obtains
\begin{equation}
\langle Q_jQ_{j^\prime}\rangle=
\delta_{jj^\prime}\frac{\hbar}{2\omega_j}
\coth\left(\frac{\hbar\omega_j}{2k_BT}\right).
\end{equation}
Substitution of  this result into (\ref{q2}) gives 
\begin{equation}
\langle {q_i}^2\rangle=\frac{\hbar}{m\omega_0}\frac{1}{\pi}
\int_0^{\frac{\pi}{2}}\frac{\sin^2\left(2i\theta\right)}{\sin\theta}
\coth\left(
\frac{\hbar\omega_0\sin\theta}{k_BT}\right)d\theta.
\end{equation}
In the high temperature limit, $\hbar\omega_0/k_BT\ll 1$, one can use the 
approximation
$\coth(x)\approx 1/x$, which leads to the classical result (\ref{msd}). 

For  the ultimate quantum case $T=0$,  
$\coth(x)$ goes to one, and the MSD takes the form
\begin{equation}
\langle{q_i}^2\rangle=\frac{\hbar}{m\omega_0}\frac{1}{\pi}\int_0^{\frac{\pi}{2}}\frac{\sin^2\left(2i\theta\right)}{\sin\theta}d\theta
\end{equation}
where  the integral increases with $i$ logarithmically 
\begin{equation}
\int_0^{\frac{\pi}{2}}\frac{\sin^2\left(2i\theta\right)}{\sin\theta}d\theta\sim \frac{1}{2}\ln{\sqrt{i}}.
\end{equation}
Thus the MSD 
due to quantum zero-point fluctuations
reads as follows,
\begin{equation}
\langle{q_i}^2\rangle\sim\frac{\hbar}{2\pi m\omega_0}\,\ln{i}.
\end{equation}
For $\omega_0\sim 10^{12}\,s^{-1}$, $m\sim 10^{-27}\,kg$ (proton),
and $\ln i\sim 1$, 
the above equation gives  
$\langle q_i^2\rangle^{1/2}\sim 1\AA$.

\section{Dynamics}
With the delocalized  character of atomic motion 
in a long chain established, let us consider the  
question about the dynamics of this motion.
It can be conveniently 
described in terms of a velocity correlation function  
\begin{equation}
C(t_1,t_2)=\langle v(t_1)v(t_2)\rangle.
\end{equation}
For instance, integrating 
$C(t_1,t_2)$ one obtains 
the MSD $\langle \Delta q^2(t)\rangle
=\langle [q(t)-q(0)]^2\rangle$ of an atom:  
\begin{equation}
\langle \Delta q^2(t)\rangle=\int_0^t dt_2\int_0^t dt_1\,C(t_1,t_2).
\end{equation}  
Using stationarity of the process $v(t)$, 
$C(t_1,t_2)\equiv C(t_2-t_1)$
and integrating by parts, one gets
\begin{equation}
\langle \Delta q^2(t)\rangle=
2\int_0^t d\tau\,(t-\tau)C(\tau).
\label{msd2}
\end{equation} 
Next, one can show~\cite{hh} that 
the Laplace-Fourier transform of the velocity correlation function
$\tilde C(\omega)=\int_0^\infty dt\,e^{-i\omega t} C(t)$ determines
the dynamical mobility $\mu(\omega)$
of a charged  atom,
\begin{equation}
\mu(\omega)=\frac{e}{k_BT}\,\tilde C(\omega).
\label{mu}
\end{equation}
For an isolated chain, $C(t)$ is a Bessel function
\begin{equation}
C(t)=\frac{k_BT}{m}\,J_0(2\omega_0t).
\label{C}
\end{equation}
This result can be obtained using the normal mode transformation of the 
previous section. Indeed, since
\begin{equation}
v_i(t)=\frac{1}{\sqrt{m}}\sum_{j=1}^N A_{ij}P_j(t)
\end{equation}
and
\begin{equation}
P_j(t)=P_j(0)\cos\omega_jt-\omega_jQ_j(0)\sin\omega_j t,
\end{equation}
one obtains for the correlation $C_i(t)=\langle v_i(0)v_i(t)\rangle$
\begin{equation}
C_i(t)=
\frac{1}{m}\sum_{j,k=1}^NA_{ij}A_{ik}\langle P_j(0)P_k(0)\rangle\cos\omega_jt.
\end{equation}
Assuming that initial distribution of coordinates and momenta 
is canonical, one gets $\langle P_j(0)P_k(0)\rangle=\delta_{jk}k_BT$. 
Then 
\begin{equation}
C_i(t)=
\frac{k_BT}{m}\sum_{j=1}^NA_{ij}^2\cos\omega_jt.
\end{equation}
Substituting the explicit expressions for $A_{ij}$ and $\omega_j$ and converting the sum into an integral  one obtains
\begin{equation}
C_i(t)=
\frac{k_BT}{m}\,\frac{4}{\pi}\int_0^{\pi/2} d\theta\, \sin(2i\theta)\,
\cos(2\omega_0\sin\theta),
\end{equation} 
or 
\begin{equation}
C_i(t)=
\frac{k_BT}{m}\,\Bigl\{
J_0(2\omega_0t)-J_{4i}(2\omega_0t)
\Bigr\}.
\end{equation} 
For large $i$ the term with $J_{4i}$ may be neglected , 
and one recovers the result 
(\ref{C}).

Using (\ref{msd2}) and (\ref{C}) one obtains for the MSD
\begin{equation}
\langle \Delta q^2(t)\rangle=2D_at-2D_at\,J_1(2\omega_0t),
\label{msd3}
\end{equation} 
were the diffusion coefficient is
\begin{equation}
D_a=\int_0^\infty dt\,C(t)=\frac{k_BT}{2m\omega_0}.
\end{equation} 
For long time, $\omega_0t\gg 1$, the first  term in (\ref{msd3})
dominates, so that the atomic motion is diffusive,
$\langle \Delta q^2(t)\rangle\sim 2D_at$. 
This result was first 
discussed by Rubin~\cite{Ru}. 
In the  same limit 
the mobility is purely real and does not depend on frequency, 
$\mu(\omega)\approx e/2\omega_0 m$.

\section{Dissipative chains} 
In previous sections the interaction of chains has been  
neglected which is 
very likely to be an over-simplification.
The problem of dynamics of interacting  chains  may be considered as  
a generalization of 
the Frenkel-Kontorova model about a chain 
in an external spatially periodic potential field~\cite{Gillan}.
In quasi-1D systems  this periodic potential is created by adjacent chains 
and is not static, which makes the problem very difficult~\cite{Savin}.
It was suggested in \cite{hh} that some insight can be achieved by 
modelling the chains interaction using  the Langevin approach. 
Namely, one may assume that the force exerted on an atom by adjacent chains 
can be written as the sum of a regular dissipative force 
linear in the atom's velocity, $-\gamma \dot q_i$, and a 
fluctuating term $\xi_i(t)$. With this assumption
the equation of motion of the atoms takes the form
\begin{equation}
m\ddot{q}_i(t)=k(q_{i-1}+q_{i+1}-2q_i)-\gamma\dot q_i(t)+\xi_i(t).
\end{equation}
Let us treat the fluctuating term $\xi_i(t)$ as a zero centered white noise
which is not correlated for different atoms and related to the friction constant $\gamma$ through the conventional fluctuation-dissipation relation,
\begin{equation}
\langle \xi(t)\rangle=0,\,\,\,
\langle \xi_i(0)\xi_k(t)\rangle=2k_BT\gamma\,\delta_{ik}\,\delta(t).
\end{equation}
Such approach is common in polymer physics to describe the polymer-solvent 
interaction. In that case the friction constant $\gamma$ can be expressed
in terms of the  solvent viscosity. In our model, the dissipation and 
fluctuating terms describes interaction with other chains, and $\gamma$
is an adjustable parameter.

Let us find the velocity correlation function of an atom for the model of 
dissipative chains. Referring again to the normal mode transformation
(\ref{transformation}), the inverse has the form 
\begin{equation}
Q_j=\sqrt{m}\sum_{i=1}^N A_{ij}q_i,\,\,\,
P_j=\frac{1}{\sqrt{m}}\sum_{i=1}^NA_{ij}p_i.
\end{equation}
Then from the above equation of motion one finds the equation for $Q_j(t)$
\begin{equation}
\ddot{Q}_j(t)=-\omega_j^2Q_j(t)-\lambda \dot{Q}_j(t)+\eta_j(t).
\label{EMNM}
\end{equation}
Here
\begin{equation}
\eta_j(t)=\frac{1}{\sqrt{m}}\sum_{i=1}A_{ij}\xi_i(t)
\end{equation}
has a meaning of the random force for a normal mode $Q_j$, and
$\lambda=\gamma/m$ is the inverse velocity relaxation time for an atom.
Using the method of Laplace transform
the solution of (\ref{EMNM}) can be written as
\begin{equation}
Q_j(t)=a_j(t)Q_j(0)+b_j(t)\dot{Q}_j(0)+\int_0^td\tau
b_j(t-\tau)\eta_j(\tau).
\end{equation}           
The explicit form of the functions  $a_j(t)$ and $b_j(t)$ depends on
the sign of the difference $\omega_j-\lambda/2$. We consider here only the case of an  over-damped chain when $\lambda>2\omega_0$ (and therefore  
$\lambda>2\omega_j$ for any 
normal mode $j$). In this case 
\begin{eqnarray}
\,\,\,\,\,\,\,\,\,\,\,\,
a_j(t)&=&e^{-\frac{\lambda}{2}t}\left\{\frac{\lambda}{2\Omega_j}
\sinh\Omega_jt+\cosh\Omega_jt\right\},\\
\,\,\,\,\,\,\,\,\,\,\,\,
b_j(t)&=&\frac{1}{\Omega_j}e^{-\frac{\lambda}{2}t}\sinh\Omega_jt,
\end{eqnarray}
where $\Omega_j=\sqrt{\lambda^2/4-\omega_j^2}$. Then for the 
velocity correlation function $C(t)=\langle \dot{q}_i(t)\dot{q}_i(0)\rangle$
one gets
\begin{equation}
C(t)=\frac{k_BT}{m}\sum_jA_{ij}^2\,\dot{b}_j(t).
\label{aux33}
\end{equation}
For the very strong dumping $\lambda\gg\omega_0$, $\dot b_j(t)$
can be approximated as 
\begin{equation}
\dot b_j(t)=
exp(-\lambda t)-\left(\omega_j/\lambda\right)^2
exp\left(-\omega_j^2t/\lambda\right).
\end{equation}
Transforming the sum (\ref{aux33}) into an integral
one obtains the velocity correlation function in the form
\begin{equation}
C(t)=\frac{k_BT}{m}e^{-\lambda t}+
\frac{k_BT}{m\lambda}\,\,\frac{d}{dt}\left\{
e^{-\alpha t}\,\,I_0(\alpha t)
\right\}
\label{C2}
\end{equation}
where $I_0(x)$ is the modified Bessel function, and 
$\alpha=2\omega_0^2/\lambda$. Note that $\alpha\ll\lambda$.

Substituting the above expression for $C(t)$ into (\ref{msd2})
one obtains
\begin{equation}
\langle \Delta q^2(t)\rangle=\frac{2k_BT}{\gamma}\,te^{\alpha t}
\Big\{I_0(\alpha t)+I_1(\alpha t)\Big\}.
\label{MSD3}
\end{equation}
For long time, $t\gg\alpha^{-1}$, this expression gives sub-diffusive behavior
\begin{equation}
\langle \Delta q^2(t)\rangle\sim 2F\sqrt{t}
\end{equation}
with the mobility factor
\begin{equation}
F=\frac{k_BT}{\omega_0\sqrt{\pi\gamma m}}.  
\end{equation}
Sub-diffusive motion with the MSD growing as $\sqrt{t}$ 
is characteristic feature for any
form of the single file diffusion when overdamped Brownian particles
constrained  to move in one dimension and are not allowed to pass each 
other~\cite{Lev}. Its consequence for the hitchhiking model is that
the diffusion constant depends on the inter-chain transition 
rate $W_\perp$,
as follows from  equation (\ref{Destim}), 
$D_\parallel=F\sqrt{W_\perp}$. Note however, that this result 
holds only for the 
transport on the time scale much longer than $1/W_\perp$; 
it does not apply for short 
chains when inter-chain transitions are negligible.


Another
consequence of sub-diffusive transport is that, in contrast to the approximation of non-interacting  
chains, 
the dynamical mobility essentially depends on frequency.
Using (\ref{mu}) one can find that for 
$\omega\ll\alpha$ both real and imaginary parts increases with 
$\omega$ as $\omega^{1/2}$, namely  
\begin{equation}
\mu(\omega)=\frac{e}{2\omega_0}\sqrt{\frac{\omega}{2m\gamma}}(1-i).
\end{equation}
Power frequency dependence $\mu(\omega)\sim \omega^s$ with $0<s<1$ is 
typical for many disordered systems.  Note however, that for frequency 
lower than the inter-chain transition rate, one has to take into account
inter-chain hopping of carriers. On this time scale the carrier 
diffusion is normal
$\langle \Delta q^2(t)\rangle\sim t$, and $\mu(\omega)$ is almost 
frequently independent.

\section{Concluding remarks}
This paper promotes the idea that in quasi-1D systems long-range structural 
fluctuations may carry localized electrons over a considerable distance  
and that for sufficiently low temperature this may be the dominating 
mechanism of charge transport. Most of the results are obtained under 
very idealized assumptions  and can hardly be regarded as anything but   
toy-model calculations. 
On the other hand, qualitative 
predictions of the model seem quite general. For instance,  while 
the character of 
time dependence for the atomic MSD may depend on 
many factors, the  
linear dependence of the MSD on temperature  
is a general property, which is responsible for a  
temperature independent  hitchhiking mobility.  

The assumption of  one-dimensional
dynamics of the chains is also not essential.  In this paper 
we assumed that the chains are parallel and form 
strongly anisotropic 2D or 3D crystal. One may argue that
the model of hitchhiking electronic transport may  be relevant to 
polymer systems, where monomers move sub-diffusively 
in three dimensions~\cite{polymer}.  Another generalization is 
the case when the electronic localization length is much larger than 
the lattice spacing, $L\gg a$. This problem is related to the  
dynamics of a large cluster~\cite{cluster} 
rather than of a single particle in a chain.

\vspace{0.3cm}

\section{acknowledgement}
The work was supported by a grant from NSERC.

\end{document}